\documentclass[aps,twocolumn,showpacs,prb,10pt]{revtex4-1}

\usepackage{amsmath}
\usepackage{amssymb}
\usepackage{epsfig}

\begin{document}

\title{Large tunable photonic band gaps in nanostructured doped semiconductors}
\author{ J. Leon$^1$, T. Taliercio$^2$}
\affiliation{(1) Laboratoire de Physique Th\'eorique et Astroparticules
CNRS-IN2P3-UMR5207, \\
and (2) Institut d'Electronique du Sud, CNRS-INSIS-UMR5214,\\
Universit\'e Montpellier 2, 34095 Montpellier (France)}

\begin{abstract}
A plasmonic nanostructure conceived with periodic layers of a doped
semiconductor and passive semiconductor is shown to generate spontaneously
surface plasmon polaritons thanks to its periodic nature. The nanostructure is
demonstrated to behave as an effective material modeled by a simple dielectric
function of ionic-crystal type, and possesses a fully tunable photonic band gap,
with widths exceeding 50\%, in the region extending from mid-infra-red to
Tera-Hertz.
\end{abstract}

\pacs{42.25.Bs, 42.79.-e, 78.20.Ci \hfill Phys. Rev. B {\bf 82} (2010) 195301}

\maketitle

\section{Introduction}

The science of surface plasma waves on metallic-dielectric interfaces
\cite{Raether}, has received such interest as to become a true branch of
physics, the palsmonics, and opened new perspectives in the control of
light-matter interactions, see e.g. Refs.\cite{Zayatsa,Maier,Bliokh}. The
surface plasmon polariton (SPP), which results from coupling electromagnetic
field with collective oscillations of electrons supported by the
metal/dielectric interface, has unique physical properties based on enhanced
nanolocalized optical fields, allowed by the negative dielectric constant of the
metal below the plasma frequency. For example SPP propagation can be controlled
by waveguides \cite{Gramotnev} or by plasmonic crystals whose metal film
periodically nanostructured induces the plasmonic band gap
\cite{Kitson,Bozhevolnyi}. 

Engineering of surface plasmons using nanostructuration  made it possible to
develop a new range of materials with remarkable optical properties such as
extraordinary optical transmission \cite{Ebbesen}, optical filtering
\cite{Collin}, optical magnetism \cite{Temnov}, second harmonic generation
\cite{Pu} or higher harmonic generation in extreme UV \cite{Kim}, and also to
control the optical processes at the femto-second scale \cite{Kekatpure,Utikal}.
The main concern is to control the optical properties associated with surface
plasmons by engineering effective materials through periodicity,
size and metal shape of the nano-objects. For instance a recent theoretical
study proposed the nanosturcturation of metallic nano-sticks along the three
direction of space to highlight a strong coupling between the incident light and
the free electrons of the metal which opens a photonic band gap (with a stop
band of 6.3\%) in the telecom wavelength range  \cite{Huang}.

The main idea is to ensure opening of the gap by using the polaritonic nature of
surface plasmons rather than Bragg reflections. As a matter of fact, the
dispersion relation of an electromagnetic wave propagating along the interface
between a metal and a dielectric, the so-called surface wave, is similar to the
one resulting from a strong coupling between a photon and an oscillator, it is
said to be ionic-crystal-like \cite{Raether}. However, while the low
energy surface plasmon branch is purely two-dimensional, the high energy
branch of the dispersion relation is radiative and therefore a photon
propagating along the interface will not see the stop band. Studies of such
systems dates back to 1969 \cite{Economou}, revealing three branches in the SPP
dispersion relation, two of them corresponding to the anti-symmetric modes, the
third one to the symmetric modes. 

We study light propagation in a periodic nanostructure constituted of planar
layers of a metal and a dielectric, sketched in Fig.\ref{fig:1}, assumed
infinitely periodic and infinite in the second transverse direction. We
demonstrate that this metal-dielectric nanostructure, worked as a planar
waveguide array, but outside the guided mode regime, naturally and spontaneously
generates SPP and then behaves as an effective three-dimensional material whose
optical properties are completely and precisely understood by means of a single
effective dielectric function of the ionic-crystal type, the formula
(\ref{epsilon-eff}) below.

Moreover, using, instead of a metal, a doped semiconductor is a means, by
adapting the doping level, to set the plasma frequency in a chosen range, such
as, in particular, to avoid absorption due to interband transitions. Then the
simplicity of the obtained effective dielectric function allows one to determine
readily how the choice of the materials and the dimensions of the nanostructure
will tune the width of the stop band, and how the doping level will tune its
position. The properties of the effective material in the gap region is the
principal concern of our study with future interest in its switching properties
in nonlinear regimes, see e.g. Refs.\cite{Porto,Wurtz}.

We thus demonstrate the polariton nature of the system when it is worked with
transverse magnetic (TM) fields, in the long-wave approximation. Similar studies
in literature focus mainly on the SPP properties of metal-dielectric structures,
looking thus at the lower branch of the dispersion curve. Our approach predicts
the behavior of the system on the entire spectrum, namely the lower polariton
branch, the stop band and the upper polariton branch. Last we show that the
approach applies as well for transverse electric (TE) fields. The structure
works then as a simple metal for which we compute explicitly the high frequency
dielectric constant and the (tunable) plasma effective frequency.

\section{Main result}

 From now on, all frequencies are normalized to the plasma frequency $\omega_p$
the wave numbers to $k_p=\omega_p/c$, the lengths to $k_p^{-1}$ (including
spatial variables) and time to $\omega_p^{-1}$. Moreover  we name frequency the
actual angular frequency, units $rad\cdot s^{-1}$, and all dielectric constants
are the relative ones. The doped semiconductor working as a metal, we use a
Drude model with optical index $n(\omega)$ and thus the final index profile
$n(z,\omega)$ of the periodic array can be written in the elementary cell as
\begin{equation}\label{n2z}
n^2(z,\omega)=\left\{\begin{array}{lr}
\varepsilon_{_1}, & z\in[-b,0],\\
\varepsilon\Big(1-\dfrac{1}{\omega(\omega+i\gamma)}\Big),
& z\in[0,a] . \end{array}\right.
\end{equation}
The damping coefficient $\gamma$ is also dimensionless and
scaled to $\omega_p$. 
\begin{figure}[ht] 
\centerline{\epsfig{file=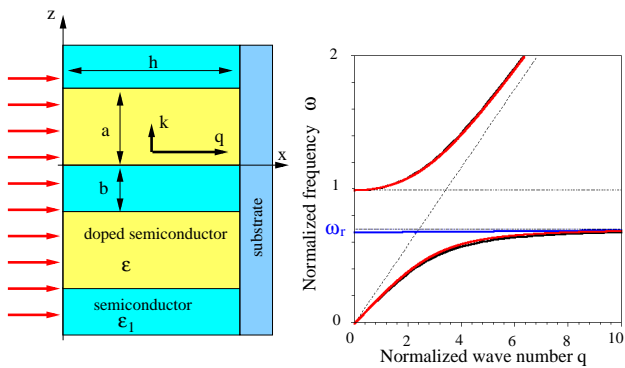,width=\linewidth}}
\caption{(color online) LEFT: scheme of the nanostructure, infinitely periodic
in direction $z$ and infinite in direction $y$. RIGHT: example of the SPP
dispersion law $\omega(q)$. The parameter values are given in Eq.(\ref{param}).
The red and blue curves are the true dispersion relation solutions of Eqs.
(\ref{red}) and (\ref{blue}) respectively, the black curves (almost
indistinguishable from the red curves) are the plot of the ionic crystal
dispersion law (\ref{w-approx}). The dashed line is the light line in the
passive semiconductor.}\label{fig:1}
\end{figure}

We shall demonstrate that, under TM irradiation, in the
long wave limit, the resulting metamaterial is equivalent to a single layer
having the following effective dielectric function 
\begin{equation}\label{epsilon-eff}
 \varepsilon_{_{\rm eff}}=\tilde\varepsilon
 \,\frac{\omega(\omega+i\gamma)-1}{\omega(\omega+i\gamma)-\omega_r^2},
\end{equation}
with the resonant dimensionless eigenfrequency $\omega_r$ and
high frequency dielectric constant $\tilde\varepsilon$ given by
\begin{equation}\label{omegar}
 \omega_r^2=\frac{b\varepsilon}{a\varepsilon_{_1}+b\varepsilon},
 \quad \tilde\varepsilon=
\frac{(a+b)\varepsilon_{_1}\varepsilon}{a\varepsilon_{_1}+b\varepsilon}.
\end{equation}
Eq.(\ref{epsilon-eff}) is identical to the dielectric function of an ionic
crystal with relative high frequency dielectric constant $\tilde\varepsilon$,
characteristic transverse frequency $\omega_r$, and longitudinal frequency
$\omega_p=1$ (by normalization). As a matter of fact, Fig.\ref{fig:1} displays
an example of the exact dispersion relation for the anti-symmetric modes (red
curves)  compared to the ionic crystal dispersion law
$\omega=q/\sqrt{\varepsilon_{_{\rm eff}}}$ (plotted with $\gamma=0$ as black
curves). The blue curve corresponds to the symmetric modes as explained in
Sec.\ref{sec:polariton}.

\section{Simulations}

We perform numerical simulations of a realistic system submitted to TM normal
incidence and calculate the reflection and transmission coefficients. The
presented spectra are simulated with the commercial software package running
finite difference time domain \cite{FDTD}. We shall illustrate our task by using
InAs for both semiconductors (high-frequency dielectric constants
$\varepsilon_{_1}=\varepsilon=11.7$), doping with silicon (Si) at a density of
$10^{20}$ that produces a plasma frequency $\omega_p=3.42\,10^{14}\,rad/s$
and hence $k_p=1.14\,rad/\mu m$ \cite{plasma}. The dimensions of the
nanostructure and the damping factor are taken as
\begin{equation}\label{param}
 a=b=0.2\,{\rm \mu m},\quad  h=1\,{\rm \mu m},\quad \gamma=10^{13}\,rad/s,
\end{equation}
which means normalized quantities $a=b=0.23$, $h=1.14$ and $\gamma=0.03$.
\begin{figure}[ht]
\centerline{\epsfig{file=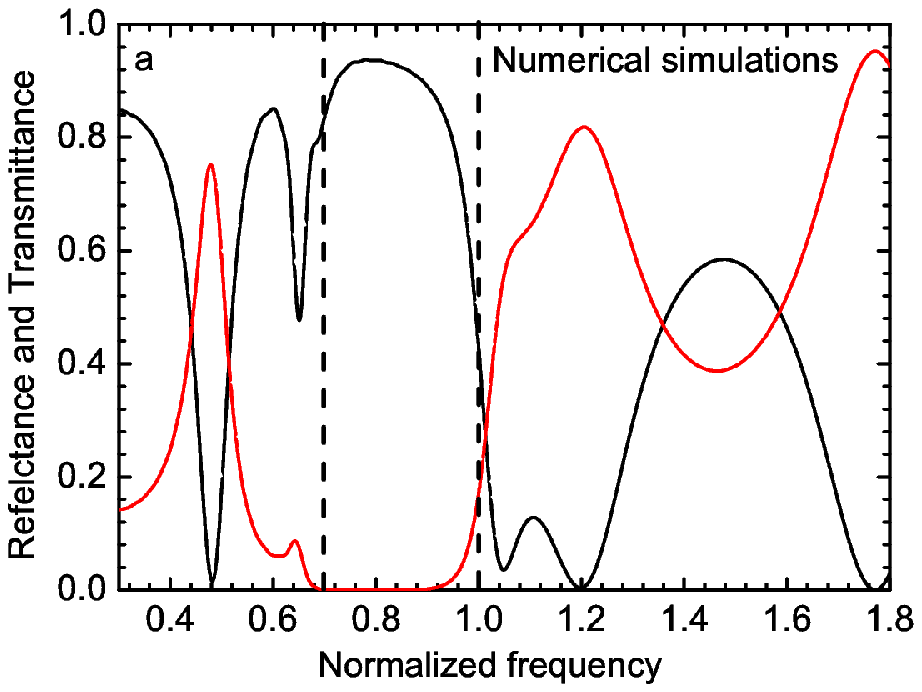,width=0.8\linewidth}}
\centerline{\epsfig{file=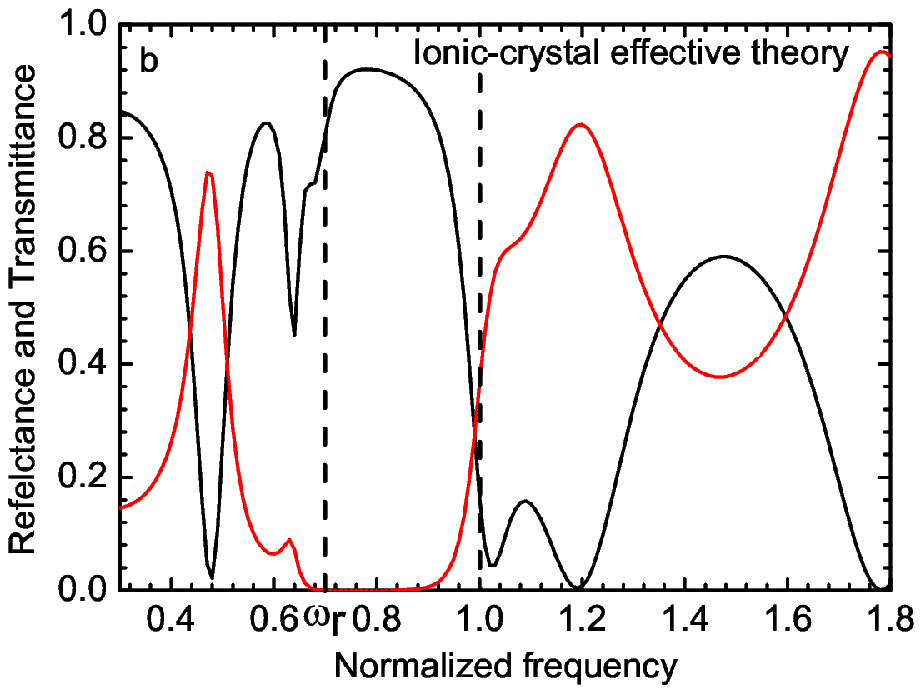,width=0.8\linewidth}}
\caption{(color online) Plots of the reflectance (black curve) and transmittance
(red curve) obtained (a) from simulations of the structure at normal incidence.
and (b) by the effective dielectric function (\ref{epsilon-eff}) with formulas
(\ref{ref-trans}). The resonant frequency $\omega_r$ given in (\ref{omegar})
defines the stop band $[\omega_r,1]$.}\label{fig:2}
\end{figure}

Fig.\ref{fig:2} shows the result of a simulation where transmittance and
reflectance are plotted in terms of the frequency of the incident TM plane wave
at normal incidence. It is then quite remarkable that the spectra are
almost identical, which evidences our main claim, though Fig.\ref{fig:2}b has
been obtained with two major approximations: the ionic-crystal dielectric
function (\ref{epsilon-eff}) and the effective layer modelization. 

The obtained normalized stop band $[\omega_r,1]$ can then be easily broadened by
playing with the dimensions of the nanostructure,  the plasma frequency
$\omega_p$, and the materials. For example changing the passive semiconductor
from InAs to GaSb (dielectric constant $\varepsilon_{_1}=14.4$) increases the
stop band from 29\% to 33\% for a depth $h=1\,{\rm \mu m}$. Last, in order to
confirm that the stop band is not the result of  a Bragg scattering, we made a
series of simulations with an incident angle varying from 0 to 15 degrees, and
with layer thicknesses up to $a=b=0.7\,{\rm \mu m}$, without any noticeable
change in the spectra.

\begin{figure}[ht]
\centerline{\epsfig{file=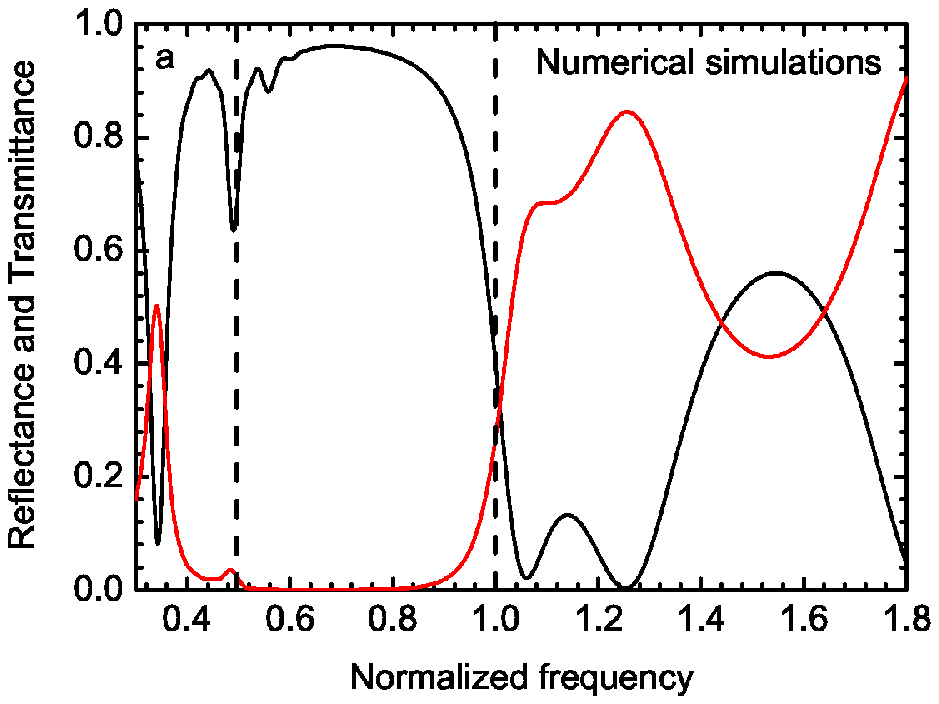,width=0.8\linewidth}}
\centerline{\epsfig{file=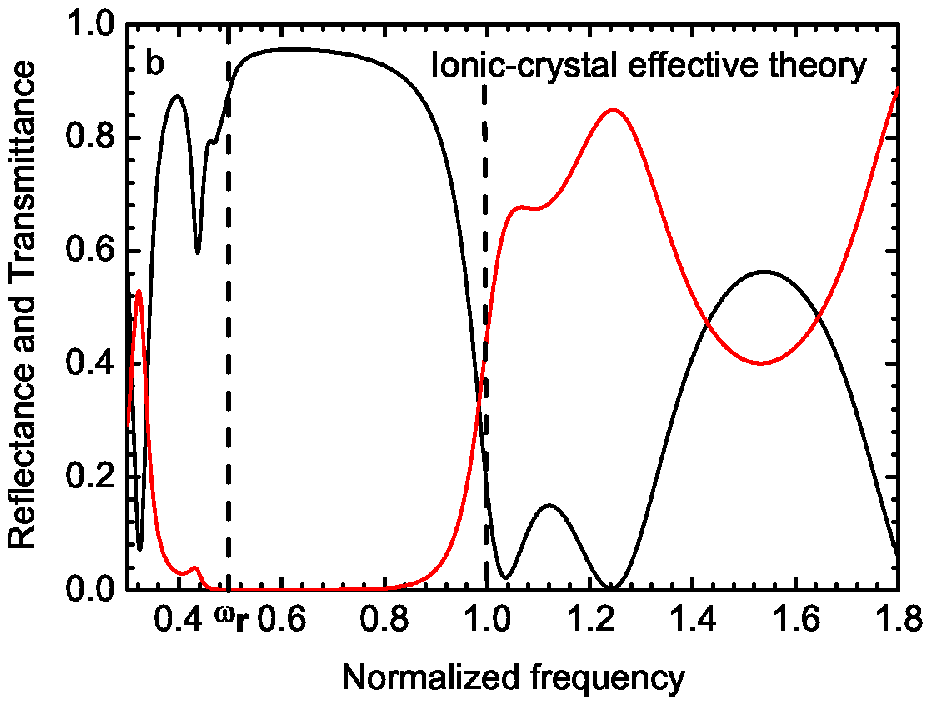,width=0.8\linewidth}}
\caption{(color online) Reflectance (black curve) and transmittance
(red curve) obtained as in Fig.\ref{fig:2} now with $a=0.3\,\mu\,m$
and $b=0.1\,\mu\,m$, the other parameters being the same.}\label{fig:3}
\end{figure}
Another example is displayed in Fig.\ref{fig:3} where we have modified only the
geometry of the structure by increasing the width $a$ of the doped semiconductor
to $0.3\,\mu\,m$ and reducing the width $b$ of the passive semiconductor to
$0.1\,\mu\,m$. The size of the stop band then reaches 50\% of $\omega_p$. Small
discrepancies appear in the lower end of the spectrum, related to the long-wave
approximation which, in the cases $a\ne b$, is less accurate. We found that the
lower branch of the dispersion relation (\ref{w-approx}) is for $a\ne b$ less
close to the actual branch given by the solution of (\ref{red}).

\begin{figure}[ht]
\centerline{\epsfig{file=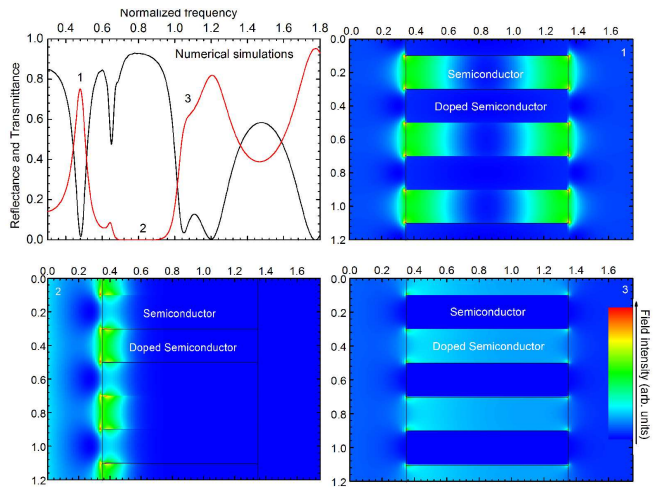,width=\linewidth}}
\caption{(color online) Field intensity plots for the parameter
values used in Fig.\ref{fig:1}, reproduced in the first graph
where numbers indicate the 3 values of the frequencies which then
index the corresponding intensity plots.}\label{fig:4}
\end{figure}

It is now worth understanding the field intensity distribution for some
representative points of the spectrum. In the lower branch of the dispersion
curve, the graph (1) of Fig.\ref{fig:4} shows that transmission is accomplished
by means of surface plasmons (in that case both $k$ and $k_1$ are pure imaginary
numbers) and that the peak of transmission is due to a stationary mode along
$Ox$ pinned at the interfaces and living essentially inside the passive
semiconductor. The next graph (2) is plotted for a frequency inside the stop
band and shows a field exponentially decreasing along $Ox$, as indeed $q$ is now
a pure imaginary number. The last graph (3) is taken for a frequency in the
upper branch of the dispersion curve. It shows a transmission essentially inside
the doped semiconductor (again with a stationary mode along $Ox$). This
transmission can be understood by noting that, at this frequency, the index of
the doped semiconductor is smaller than the index $\varepsilon_{_1}$ of the
passive semiconductor.

With all these simulations in hand, it is time to demonstrate now the essential
result, namely to derive the expression of the effective dielectric function
(\ref{epsilon-eff}). This is done in the following section simply by using
Maxwell's equations and continuity conditions for an infinitely periodic
structure.

\section{TM field: the polariton machine}\label{sec:polariton}

We thus calculate the dispersion relations and the resulting reflection
coefficient by assuming the nanostructure to act as a single layer with an
effective dielectric function. This is done to show first that the optical
properties of the metamaterial indeed result from SPP generation. Second, it
allows us to study the long wave limit and demonstrate that the dielectric
function has the explicit simple approximate expression (\ref{epsilon-eff}). 

Considering an infinitely periodic nanostructure, we may extract the elementary
cell $z\in[-b,a]$ and connect the interfaces $z=-b$ to $z=a$ by usual continuity
relations. As we expect to excite surface waves at the interfaces, we
first consider TM field and seek solutions
\begin{equation}
{\bf E}(x,z,t)=\left(\begin{array}{c}
E(z)\\ 0 \\ F(z) \end{array}\right) e^{i (\omega t-q x)},
\end{equation}
as the continuity of the tangential components readily implies that the
$x$-component $q$ of the wave vector assumes the same values in either region.
In the Drude model, the Maxwell equation reduces to the usual Helmoltz
equation for $E(z)$, together with an explicit expression for $F(z)$, with the
optical index function (\ref{n2z}), namely
\begin{align}
&\frac{\partial^2 E}{\partial z^2}+[n^2(z,\omega)\,\omega^2-q^2]E=0,
\label{Helm}\\
& [n^2(z,\omega)\,\omega^2-q^2]F=-iq\frac{\partial E}{\partial z}.\label{E-F}
\end{align}
According to Eq.(\ref{n2z}) we may define the
eigenvalues $k_1$ and $k$ in each region by
\begin{align}\label{k1k-bis}
& z\in[-b,0]\ :\quad && k_1^2=\varepsilon_{_1}\omega^2-q^2,
\nonumber\\
& z\in[0,a]\ :\quad && k^2=\varepsilon\omega^2(1-\frac{1}{\omega\omega'})-q^2,
\end{align}
where we define $\omega'=\omega+i\gamma$. Solutions to (\ref{Helm}) are
then sought under the form
\begin{align}\label{elec-field}
& z\in[-b,0] && E_1(z)=A_1e^{ik_1z}+B_1e^{-ik_1z},
\nonumber\\
& z\in[0,a] && E(z)=Ae^{ikz}+Be^{-ikz}.
\end{align} 
Continuity relations in $z=0$ and $z=a,\,-b$ for $E(z)$ and $F(z)$  then result
in an algebraic homogeneous linear system whose solvability
condition produces a $4\times4$ determinant. It actually factorizes in two
terms which can be directly retrieved by assuming first the sub-case
$E(a)=-E(0)$, which gives
\begin{equation}\label{red}
\varepsilon_{_1} k\,\tan(ak/2)+ 
\varepsilon k_1(1-\frac{1}{\omega\omega'})\,\tan(bk_1/2)=0.
\end{equation}
This case is called anti-symmetric because, due to relation (\ref{E-F}), the
transverse field component $F(z)$ is anti-symmetric on the elementary cell
$[-b,a]$. Then the symmetric case is obtained for  $E(a)=E(0)$ for which
\begin{equation}\label{blue}
\varepsilon k_1(1-\frac{1}{\omega\omega'})\,\tan(ak/2)+
\varepsilon_{_1} k\,\tan(bk_1/2)=0.
\end{equation}
Note that the complete electric field is neither symmetric nor anti-symmetric
because of relation (\ref{E-F}), still the dominant term is played by the
transverse component $F(z)$ which has a defined symmetry. The above two
equations must be read with the definitions (\ref{k1k-bis}) as relations between
$\omega$ and $q$. As usual these are implicit expressions solved numerically
which, for the dispersion relation $\omega(q)$, furnishes the graph of
Fig.\ref{fig:1} where the red curve corresponds to the modes (\ref{red}) and the
blue curve (almost a straight line) shows the modes obtained from
Eq.(\ref{blue}). These graphs are obtained as usual by assuming $\gamma=0$ to
deal with real-valued expressions.

We consider now the long-wave limit where $a$ and $b$, which are scaled to
$k_p^{-1}$, are small quantities such as to allow the Taylor expansions
$\tan(ak/2)\sim ak/2$ and $\tan(bk_1/2)\sim bk_1/2$. Note that such
approximations are valid in a finite wave number domain. Then the dispersion
relation (\ref{red}) written for the wave number $q$ is easily demonstrated to
become, with help of definitions (\ref{k1k-bis}),
\begin{equation}\label{w-approx}
q^2=\tilde\varepsilon\,\omega^2\,
\frac{\omega(\omega+i\gamma)-1}{\omega(\omega+i\gamma)-\omega_r^2}
=\varepsilon_{_{\rm eff}}\omega^2,
\end{equation}
with the dimensionless resonant frequency $\omega_r$ and dielectric constant
$\tilde\varepsilon$ given in (\ref{omegar}). The same treatment being applied to
relation (\ref{blue}) shows that the wave numbers $k$ and $k_1$ cancel out and
the relation simply reduces to $\omega\omega'=\omega_t^2$ where $\omega_t$ is
defined below in Eq.(\ref{omegat}), and which for $a=b$ reduces to $\omega_r$.
This explains why the blue curve of Fig.\ref{fig:1} tends to the straight line
$\omega_r$ (within the chosen wave number range).

Therefore we can state  that the nanostructure considered here behaves, at least
in the long-wave limit, as a single layer of active medium with effective
dielectric function $\varepsilon_{_{\rm eff}}$, which from (\ref{w-approx})
gives our fundamental result (\ref{epsilon-eff}). This result can be used to
evaluate reflectance and transmittance of a layer of depth $h$  as \cite{Born}
\begin{align}\label{ref-trans}
& R(\omega)=\left| \frac{r_{1}+r_{2}\exp[{2i\omega h n_{_{\rm eff}}}]}
{1+r_{1}r_{2}\exp[{2i\omega h n_{_{\rm eff}}}]}\right|^2,
\nonumber\\
& T(\omega)=\frac{n_3}{n_0}
\left| \frac{t_{1}t_{2}\exp[{i\omega h n_{_{\rm eff}}}]}
{1+r_{1}r_{2}\exp[{2i\omega h n_{_{\rm eff}}}]}\right|^2,
\end{align}
with the following definitions (Fresnel formulas) at normal incidence
\begin{align}\label{TM-fresnel}
&r_{1}=\frac{n_{_{\rm eff}}-n_0}{n_{_{\rm eff}}+n_0},\quad
r_{2}=\frac{n_3-n_{_{\rm eff}}}{n_3+n_{_{\rm eff}}},
\nonumber\\
&t_{1}=\frac{2n_0}{n_{_{\rm eff}}+n_0},\quad
t_{2}=\frac{2n_{_{\rm eff}}}{n_3+n_{_{\rm eff}}},
\end{align}
when the layer lies on a substrate with optical index $n_3$, when $n_0$ is the
index of the medium in the incident region and where $n_{_{\rm
eff}}=\sqrt{\varepsilon_{_{\rm eff}}}$ is given by the approximate expression
(\ref{epsilon-eff}). The graphs $(b)$ of Figs. \ref{fig:2} and \ref{fig:3} show
the plots of the above reflectance and transmittance (\ref{ref-trans}) with
$n_0=n_3=1$. Comparison with corresponding graphs $(a)$ shows the accuracy of
such a simple modelization. 

One may also compute from the relations (\ref{red}) and (\ref{blue}) the
dispersion laws $\omega(k)$ inside the doped semiconductor and $\omega(k_1)$
outside. In the long-wave approximation, one obtains that $k$ is a pure
imaginary number outside the stop band $[\omega_r,1]$ while $k_1$ is a pure
imaginary number below the resonant frequency $\omega_r$. This is why the lower
branch of the dispersion relation $\omega(q)<\omega_r$ represents a surface
plasmon: the field is exponentially located at the interfaces. 

\section{TE field: a simple metal}

It is interesting to consider the case of incident transverse electric (TE)
fields for which we demonstrate now that the system behaves as an effective
metal with a Drude-like dielectric function. Therefore the system illuminated
with TE fields does not give rise to polaritons, i.e. does not behave as a
ionic-crystal, which is related to the fact that generation of surface waves on
a metal-dielectric interface occurs only with TM fields.

The Maxwell equation  for the following TE field structure (again the wave
number $q$ in the $x$-direction is the same in both regions by continuity)
\begin{equation}\label{TE-elec-field}
{\bf E}_1=\left( \begin{array}{c} 0 \\E_1(z)\\0\end{array}\right)
e^{i(\omega t-qx)},\quad
{\bf E}=\left( \begin{array}{c}0\\E(z)\\0\end{array}\right)
e^{i(\omega t-qx)},
\end{equation}
reduces to the Helmoltz equation (\ref{Helm}). In that case the set of
continuity relations has to be completed with the magnetic field calculated
by means of $\nabla\!{\times}\!{\bf E}=-\partial_t{\bf B}$, namely
\begin{equation}\label{TE-mag-field}
{\bf B}=\frac{1}{\omega}
\left( \begin{array}{c}-i\partial_zE \\ 0\\ qE\end{array}\right)
e^{i(\omega t-qx)},
\end{equation}
and similarly for ${\bf B}_1$.
As in the preceding section, a solution is sought under the expressions
(\ref{elec-field}) but now the two sets of continuity relations at each
interface are those obtained from $({\bf E}-{\bf E}_{1})\times{\bf n}=0$ and
from $({\bf B}-{\bf B}_1)\times {\bf n}=0$, where ${\bf n}$ is the unit vector
in the direction $Oz$ normal to the interfaces. These sets are thus
\begin{align}\label{TE-cont-E}
& E(0)=E_1(0), && E(a)=E_1(-b),\nonumber\\
& \partial_z E(0)=\partial_zE_1(0), 
&& \partial_zE(a)=\partial_zE_1(-b).
\end{align}
It is then a simple task to express the above relations and, as before,
consider separately the two sub-cases $E(a)=E(0)$ and $E(a)=-E(0)$ for which we
obtain respectively  
\begin{align}
& k\,\tan(ak/2)+k_1\,\tan(bk_1/2)=0,\label{TE-red}\\
&  k_1\,\tan(ak/2)+k\,\tan(bk_1/2)=0.\label{TE-blue}
\end{align}
The wave vectors $k$ and $k_1$ are defined in Eq. (\ref{k1k-bis}), therefore
the above relation furnish the dispersion relation $\omega(q)$ in each case.

\begin{figure}[ht]
\centerline{\epsfig{file=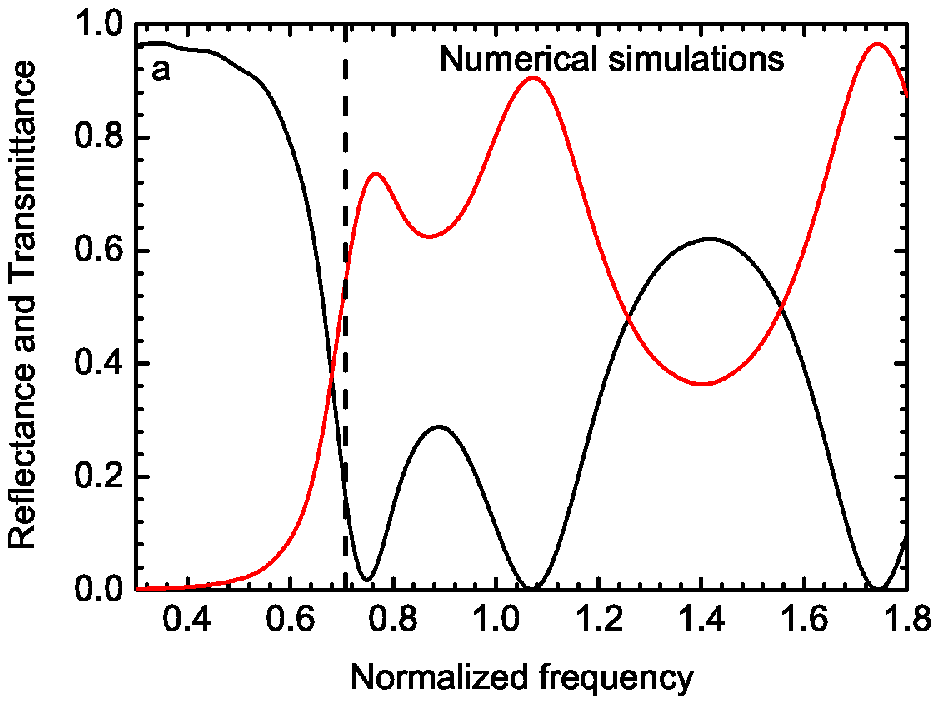,width=0.8\linewidth}}
\centerline{\epsfig{file=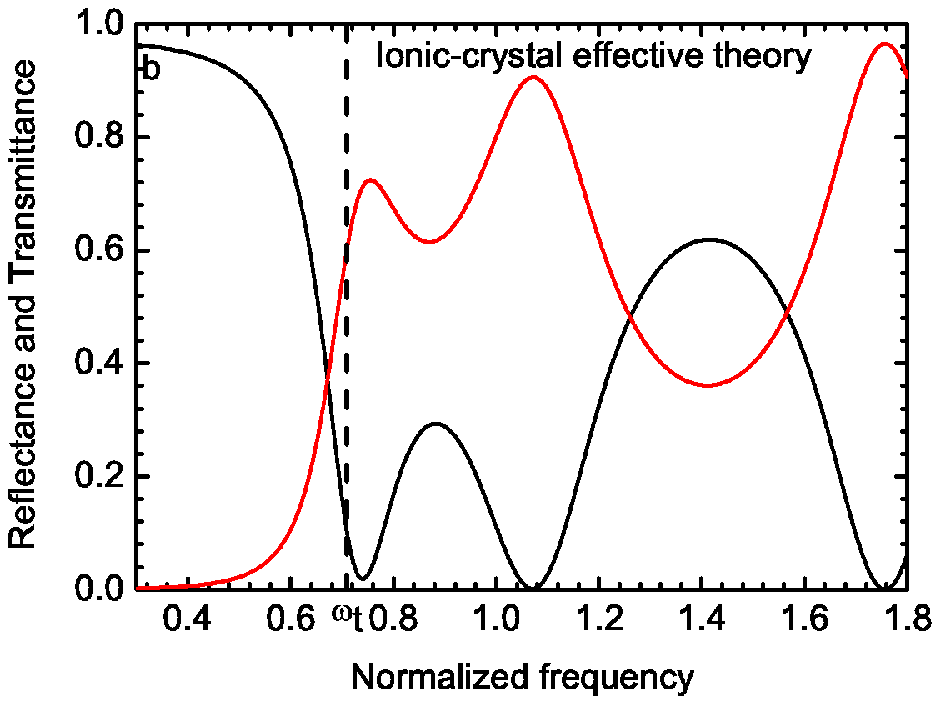,width=0.8\linewidth}}
\caption{(color online) Reflectance (black curve) and transmittance
(red curve) obtained with $a=b=0.2\,\mu\,m$ for a TE incident field with 
parameters (\ref{param}). The structure indeed works as a metallic
effective layer with the effective dielectric function defined in
Eq.(\ref{TE-eps-eff}).}\label{fig:5}
\end{figure}

We consider then the long-wave approximation for which the system is seen as a
single layer, and for which the resulting effective dielectric function can be
explicitly computed. In the limit $\tan(ak/2)\sim ak/2$ and $\tan(bk_1/2)\sim
bk_1/2$, the relation (\ref{TE-red})  can eventually be expressed as 
\begin{align}
& q^2=\varepsilon_{_{\rm eff}}\omega^2,\quad
 \varepsilon_{_{\rm eff}}=\varepsilon_{_\infty}
 \Big(1-\frac{\omega_t^2}{\omega(\omega+i\gamma)}\Big),
 \label{TE-eps-eff}\\
&\varepsilon_{_\infty}=\frac{a\varepsilon+b\varepsilon_{_1}}{a+b},\quad
\omega_t^2=\frac{a\varepsilon}{a\varepsilon+b\varepsilon_{_1}}.
\label{omegat}
\end{align}
It is a Drude-like dispersion relation with a new $\varepsilon_{_\infty}$ and a
new plasma frequency $\omega_t$ (normalized to $\omega_p$). Thus the effective
equivalent material behaves just as a metal for TE modes. Reflectance and
transmittance are given by formulas (\ref{ref-trans}) with, for normal
incidence only, Fresnel expressions (\ref{TM-fresnel}), and now with
$\varepsilon_{_{\rm eff}}=n_{_{\rm eff}}^2$ given in Eq. (\ref{TE-eps-eff}).
This  result is compared with numerical simulations on Fig.\ref{fig:5} with the
same striking efficiency as in the TM cases.

Note finally that the long-wave limit for the second relation (\ref{TE-blue})
reduces to the two solutions $k=0$ and $k_1=0$ associated with the trivial
dispersion laws obtained out of Eq. (\ref{k1k-bis}). Correspondingly, the TE
field in that case results to be a constant (along $z$) and therefore simply
vanishes due to the symmetry condition $E(a)=-E(0)$.

\section{Comments and Conclusion}

When the long-wave limit is not accurate enough, one might think of
solving numerically the dispersion relations (\ref{red}) and (\ref{blue})
to calculate $\varepsilon_{_{\rm eff}}$. However, beyond the long-wave
approximation, the main effect to take into account is the occurrence of
guided modes, primarily inside the undoped semiconductor. In that case one
must reconsider the problem since the method of the effective layer would
simply fail.

The presence of two types of modes makes it uneasy to define in general the stop
band which is not always $[\omega_r,1]$. By looking at the dispersion relations,
close to the long-wave limit, we found that the stop band is the interval
$[\omega_r,1]$ when $a<b$, but becomes $[\omega_t,1]$ when $a>b$. However in
that case, the interval $[\omega_r,\omega_t]$ is filled with symmetric modes
which are not excited under normal incidence, the effective stop band being then
still $[\omega_r,1]$. Note that when $a=b$, the 2 frequencies $\omega_r$ and
$\omega_t$ coincide, and the interval $[\omega_r,1]$ is a gap also at oblique 
incidence.

By comparing realistic numerical simulations with theoretical considerations, we
have shown that the nanostructure described in Fig.\ref{fig:1} works in TM
regime as a single layer with the effective dielectric function
(\ref{epsilon-eff}), representative of SPP generation. This result provides a
comprehensive understanding of the system as a whole, on the entire spectrum,
to show that it behaves as a ionic-crystal material. Moreover our result
provides a powerful tool to explore the possibilities offered by such a
metamaterial, e.g. by varying the sizes, the materials and the doping level,
with the opportunity to reach the visible range.

Note finally that, in view of experimental realization,  different techniques
can be adopted as e.g., epitaxy, deposition, etching or implantation. Although
these are well proved techniques, the resulting optical properties of the
structure will be affected by imperfections such as interface roughness,
diffusion, geometrical defects, which requires further studies.

\end{document}